\documentstyle [preprint,aps] {revtex}
\begin {document}
\draft
\title {Transverse Sound Revisited:  A Directional Probe of the A-Phase of
$UPt_3$}
\author {Brett Ellman and Louis Taillefer}

\address {Department of Physics,
McGill University, 3600 University Street,
Montr\'eal, Qu\'ebec, Canada H3A 2T8}

\author {Mario Poirier}
\address {Centre de Recherche en Physique du Solide,
D\'epartement de Physique,
Universit\'e de Sherbrooke, Sherbrooke, Qu\'ebec, Canada J1K 2R1}
\date{\today}

\maketitle

\begin {abstract}
We have measured the transverse ultrasonic attenuation, $\alpha_{q\epsilon}$,
in a well-characterized
single crystal of $UPt_3$ for propagation direction, ${\bf q}$, in the
basal plane and
polarization, $\mathbf{\epsilon}$, in and out of the plane.
Unlike previous measurements, the
sample was of sufficient purity to exhibit three well-defined phases as a
function of temperature and magnetic field.
We uncover a functional dependence for the attenuation in the (zero field) high
temperature phase (A-phase)
which is distinctly different than that in the low temperature (B)
phase.  Our data provide unique directional information on the quasiparticle
momentum distribution in this largely unstudied phase of $UPt_3$ and provide
the first direct evidence that the two zero-field phases possess different
symmetries.
\end {abstract}


Following the discovery of multiple superconducting phases in $UPt_3$
\cite{fisheretal}
numerous theoretical investigations have attempted to predict the gap
structure in each of the three phases. These calculations have been very
successful in predicting
measured phase diagrams obtained under magnetic
field, hydrostatic pressure, and uniaxial stress.
Indeed, the theoretical H-T-P phase diagram agrees with that
that observed under every experimental
scenario investigated to date\cite{joynt_big,review}.  While many
experiments agree as to the location of the phase lines, it is notable
that
there are very few direct measures of the {\it gap structure} within each
phase.
Here, "direct" means an
experimental probe that couples to the quasiparticle density in such a way
that the quasiparticle momentum distribution may be mapped.  Potential examples
include thermal conductivity\cite{ourkappa}, ultrasound
attenuation\cite{old_data}, muon spin resonance\cite{muon}, and point contact
spectroscopy\cite{point}, each of which has yielded insights into the gap
structure of the low temperature (B) phase.  Studies of the high temperature
(A) phase
have been very limited, however, largely due to the very small temperature
range
(about 60 mK) over which the A-phase exists.  Thus thermal conductivity
data, which have strongly constrained parameter space for the B-phase, are
insufficiently sensitive to the growth of the gap near $T_c$ to similarly
address the properties of the A-phase.  Point contact spectroscopy, while an
elegant probe of the gap structure, is similarly limited since the exact
temperature of the A/B phase transition was not known for the samples used.
Considering the small temperature region over which the A-phase may be
studied, it is essential to perform experiments on independently
characterized samples using an experimental probe which is sensitive to the
onset of superconductivity.  The data presented here satisfy these criteria
and constitute the most
definitive evidence for a different gap symmetry in the A vs. the B-phase.

Transverse ultrasonic attenuation is one of the most
powerful tools for investigating
the geometric properties of the superconducting gap.  The strengths of
transverse sound attenuation, $\alpha_{q\epsilon}(T)$,
lie in its strong dependence on the
magnitude of the gap and also on its highly directional
nature involving two independent vectors, the propagation direction, ${\bf q}$,
and the polarization, ${\bf{\epsilon}}$.  Ultrasonic attenuation changes very
sharply on entering the superconducting state.
The relative sensitivity of sound attenuation as compared to, for example,
the thermal conductivity, $\kappa$, may be
motivated by considering the energy dependence of the integrand of the
appropriate BCS quantities:

\begin{eqnarray}
\kappa &\sim& \int_{\Delta(T)}^{\infty} E^2 \frac{\partial f}{\partial E}
 dE\\
\alpha &\sim& \int_{\Delta(T)}^{\infty} \frac{\partial f}{\partial E} dE
\end{eqnarray}
Both $\alpha$ and $\kappa$ depend on the number of quasiparticles per unit
energy and thus on the derivative of the Fermi function, $\partial f /
\partial E$.  However, the expression for $\kappa$ has an energy prefactor
$E^2$
because thermal
conductivity depends on energy transport.  Since $\partial f /
\partial E$ is peaked at $E=0$, as the gap $\Delta (T)$ becomes non-zero
the energies contributing the most to the $\alpha$ integral are excluded and
the attenuation drops sharply (as $\alpha (T) = 2 f[\Delta (T)]$).
The integrand of $\kappa$,
however, is zero at $E=0$ and thus the thermal conductivity is a much weaker
function of $\Delta$ near $T_c$.

Transverse sound attenuation is also very
sensitive to gap anisotropy.  Indeed, it was transverse
attenuation data that first provided definitive evidence for a highly
anisotropic gap in $UPt_3$\cite{old_data}.
These measurements found a linear
dependence of the attenuation for $\alpha_{ab}$
(${\bf \epsilon} || \hat b$ and ${\bf q} || \hat a$) while $\alpha_{ac} \sim
T^3$.
Thus the measured attenuation was not only decidedly non-BCS
($\alpha_{ij} \sim  exp(-\Delta/T)$ for all i,j at low T)
but also manifestly anisotropic.
In the "dirty limit" (see below) appropriate to all experiments to date, the
analysis of Kadanoff and Falko\cite{Kad}, later extended to the resonant
impurity case by several groups\cite{resonant}, gives for the attenuation
(outside of any gapless regime near $T=0$)

\begin{equation}
\alpha_{q\epsilon} \propto \int_{0}^{\infty}
d\omega \frac{\partial f(\omega)}{\partial \omega}
\frac{1}{Im(\Sigma_0)}  \left<
p_{\epsilon}^2 p_q^2
\frac{\sqrt{\omega^2-\Delta(p)^2}}{\omega} \right>_p  ,
\end{equation}

where $\Sigma_0(\omega,T)$ is the
electronic self-energy (independent of
polarization or propagation direction), $\Delta(p)$ is the gap, $p_{\epsilon}$,
$p_{q}$ are the projections of momentum, $p$, along the appropriate directions
on
the Fermi surface,
and $<>_p$ is
an average over the Fermi surface restricted so that $\omega^2-\Delta(p)^2 >0
$.
The function $\Sigma_0$ includes the effects of resonant impurity
scattering\cite{pethick} which modify the energy-dependent electronic mean free
path.
The point is that the experimental parameters $q,\epsilon$
act solely as weights in the Fermi surface average of a function of $\Delta_p$.
Therefore, in principle, the transverse
attenuation allows one to directly map out the k-space location of nodes.
As an example, we consider the effect on $\alpha_{ac}$ of
a line node in the hexagonal basal plane of $UPt_3$ (e.g., $\Delta(k_c = 0) =
0$).
The
product of momenta $p_qp_{\epsilon}$ in Eq. 3 is then equal to $p_ap_c$, which
vanishes in the basal plane.
Thus the contribution of a line node to the attenuation $\alpha_{ac}$ is much
reduced as
compared to its contribution to
$\alpha_{ab}$.  That this is in qualitative agreement with the results of
Ref.\cite{old_data} constitutes strong evidence that there is a basal
plane line node in the B-phase of $UPt_3$.
The {\it relative} sizes of the $\alpha_{ij}$ for various model gaps
may all be qualitatively explained in like manner.
However, a caveat is
essential:  the functional form of $\alpha(T)$ may depend strongly on the form
of
$\Delta_p$ far from the node, even at relatively low temperatures.
Thus only a careful, self-consistent calculation can be confidently compared
with
experimental data.
Detailed calculations\cite{resonant} based on Eq. 1 found the aforementioned
experimental
results\cite{old_data} to be in qualitative
agreement with a gap having a line node in the basal plane.
These data, however, do not bear on the question of the gap structures of the
multiple phases in $UPt_3$ insofar as the sample used predated the growth of
high-purity, low strain crystals exhibiting the now familiar phase diagram.
We have performed transverse sound attenuation studies on a
single crystal of $UPt_3$ which is known to support three superconducting
phases in the temperature-field plane.
Most notably, our
data are unique in providing an incisive directional probe of the gap structure
in the high-T/low field A-phase.

The sample used was a single crystal rectangular prism approximately
$1 \times 1 \times 2.7 mm^3$ in size.  The sound was propagated along
the long dimension which in turn corresponds to the crystallographic a-axis.
Ultrasound was generated using 30 MHz lithium niobate transducers.
The attenuation was measured using a conventional pulse-echo interferometer
in a transmission geometry at frequencies ranging from 90 - 290 MHz.
The frequency dependence of the attenuation was as expected for a metal, a
point which will be considered in more detail below.
The sample
is the same one used by Adenwalla {\it et al.}\cite{adenwalla} in their careful
study of the phase boundaries using longitudinal sound velocity measurements.
We therefore know to high precision the location of the
phase boundaries {\it a priori}, an important advantage when studying the
functional form of the attenuation over the small temperature range covered by
the A-phase.  The $T_c$'s obtained from our data are in
excellent agreement with those obtained in Ref. \cite{adenwalla}:
$T_c^+ = 495$mK and $T_c^- = 435$ mK.

The measured attenuation for polarization in and transverse to the hexagonal
basal
plane are shown in Figs. 1 and 2, respectively.  The normal state attenuation
is a decreasing function of temperature, which is a consequence of the fact
that the electrical conductivity also falls with increasing temperature.
(For a Fermi liquid in the dirty limit\cite{normal_state},
$\alpha(T) \propto \sigma(T)$.)
The experimental technique determines
$\alpha(T)$ up to an additive constant corresponding to sound loss in cables,
bonding layers, and so forth.  We fix this parameter by demanding that
$\alpha(T=0) = 0$.  It is important to realize that there is no thermodynamic
reason that this condition must hold.
Indeed, some pertinent theories predict a
non-zero attenuation at zero temperature.  Therefore this experiment (or any
similar work of which we are aware) cannot directly address the issue of
a residual attenuation at T=0 reflecting a finite density of zero energy
excitations.  (This question is addressed
by recent thermal conductivity
results\cite{ourkappa}).
Independent of any theoretical interpretation,
it is evident that our B-phase results
(i.e., T $<$ 440 mK) are in
agreement with those of Ref. \cite {old_data}:
$\alpha_{ab} \sim T$ and $\alpha_{ac} \sim T^n$ with $n \sim 3$.
The exact power law
dependence of the attenuation should not be taken too seriously, particularly
for $\alpha_{ac}$.  We also note that the absolute
magnitudes of our data are in
general agreement with Ref. \cite{old_data} after appropriate frequency
normalization, as discussed below.

The main results of this work are the sharp differences observed in the
attenuation of the A-phase (500 mK $>$ T $>$ 440 mK) as compared to the
B-phase.  As may be seen in Fig. 3, $\alpha_{ab}$ drops quickly with
decreasing T before
becoming essentially constant, while $\alpha_{ac}$ has a "bump"
seemingly superimposed on
the sharply falling attenuation observed in the B-phase.
Interestingly, the data of Fig. 3
show that $\alpha(T)$ attains its low temperature asymptotic functional form
for phase B within a small range $\Delta T \leq 25$ mK below $T_c^-$.
Making an analogous assumption about the normal-state/A-phase transition,
it is tempting to assume that the essentially constant $\alpha(T)$
seen in Fig. 3 may be used to predict the gap structure of the A-phase
without worrying about the small temperature range available.  Quite aside from
any other objections, however, such an assumption neglects the effect of
superconductivity on the quasiparticle-quasiparticle scattering in $UPt_3$.
Above $T_c$, quasiparticle-quasiparticle interactions lead to a conductivity
$\sigma(T) = 1/(\rho_0 + AT^2)$ where, the elastic and inelastic
contributions are the same order of magnitude at $T_c$.
On entering the superconducting state, the quasiparticle
density falls, leading to a decrease in the inelastic term, presumably within
a fairly small temperature interval below $T_c$.  The corresponding
increase
in the quasiparticle mean free path would then lead to an increase in $\alpha$
with decreasing T.
Quantitative interpretations of the A-phase
data assuming a specific gap must carefully consider this effect, particularly
since the structure and evolution of the gap at high temperatures affect the
quasiparticle density and therefore the magnitude of the inelastic scattering.
We emphasize, however, that while changes in inelastic scattering due to
superconductivity might affect the {\it size} of the observed anisotropy,
they are not expected to {\it give rise} to anisotropy which otherwise would
not be present.

Qualitatively, our data allow us to immediately conclude
that more quasiparticles exist in the A-phase than would be present if the
B-phase extended up to the same temperature.  Furthermore, it appears that
these extra excitations preferentially scatter sound when the polarization is
in the basal plane.
To see this we plot, in Fig. 4, the data of Figs. 1 and 2 normalized to the
attenuation
at either $T_c^+$ or $T_c^-$
as a function of temperature normalized to the appropriate
critical temperature.  This allows us to compare, say, the B-phase
attenuation (for either polarization) with the attenuation in the A-phase
over the same reduced temperature range.  It is evident that $\alpha_{ab}$
is much enhanced in the A-phase as compared to the B-phase.  The data for the
c-axis polarization, $\alpha_{ac}$,
however, are roughly equal in the two phases.  The latter result implies
that there is either a serendipitous combination of inelastic scattering
effects
and quasiparticle contributions far from the gap that lead to the similarity
of the A and B-phase results, or that the gap structures in the two cases
possess the same nodal weight in the a-c plane. (According to Eq. 3,
this plane contributes most strongly to the attenuation.)  The latter
explanation is puzzling in the context of popular theories\cite{review}
(such as the $E_{1g}$ scenario mentioned in the discussion of the B-phase
results) that predict a line
node perpendicular to the basal plane in the A-phase, with the orientation of
the nodal plane determined by the direction of the antiferromagnetic order
parameter.  The latter is known to form many small domains in the sample along
the three $a^*$ (k-space) directions.
Therefore, regions of the sample would possess a line node which is not
perpendicular to either $\epsilon || c$ or $q || a$.
According to Eq. 3, such nodes would contribute more quasiparticles with
favorable
momenta than the c-axis point nodes that are postulated to exist in the
B-phase, leading to an enhanced attenuation in the A-phase.  Such
speculations, of course, emphasize the need for a complete theoretical
treatment.

The data  of Fig. 3
contain precise information on the
momentum space distributions of quasiparticles, and thus on the gap structure.
As mentioned above,
such detailed information can only be extracted by comparison with complete
calculations analogous to the published B-phase computations taking
proper account of inelastic scattering.
In the absence of pertinent theoretical predictions,
we conclude with some strictly qualitative observations on sample dependence.
The effect of
anisotropy in the quasiparticle spectrum on the attenuation critically
depends on the
measurement frequency and the purity of the sample.
In the clean limit,
defined as $ql>>1$, where
$l$ is the electronic mean free path and $q$ the phonon wavevector,
energy-momentum conservation implies
that only quasiparticles with momenta very nearly perpendicular to q may
contribute to the attenuation.  Though somewhat analogous to
the effect of the $p_\epsilon p_q$ term in Eq. 1, the clean limit attenuation
is affected by a much smaller region of the Fermi surface.
In this limit, therefore,
attenuation is a very effective probe of the anisotropy of both the Fermi
surface and the gap in a superconductor.  In fact, attenuation data in this
regime was one of the most effective early techniques used in studies
of the anisotropic gaps
found in conventional superconductors\cite{BCSans}.
We have measured the frequency dependence of the normal state attenuation
in our sample and find approximate $\omega^2$ scaling, as predicted for the
hydrodynamic regime.  Deviations from perfect scaling\cite{normal_state}
indicate,
however,
that $ql$ may be as large as 0.2 at the top of our frequency range (about 300
MHz).  We may therefore roughly estimate the quasiparticle mean free path in
the normal state as $l \approx 300$ nm.
This is in accord with estimates of the mean free path from,
e.g., de Haas-van Alphen experiments\cite{dhva}.
Cleaner samples and/or higher frequencies may thus make the
clean regime accessible in the near future.

In conclusion, we have measured the attenuation of transverse ultrasound in
the A and B phases of $UPt_3$.  Our B-phase data are in agreement with
published results\cite{old_data}.
The A-phase data are qualitatively different
and display a significant directionally-dependent enhanced density of
quasiparticles as compared to the B-phase.  In combination with resonant
impurity scattering calculations which take into account the normal state
properties, these data should provide some of the first constraints on the
symmetry of the A-phase.

Acknowledgments

We would like to thank Andreas Fledderjohann for very useful conversations
and the use of his results before publication and Mario Castonguay for
technical assistance.  This work was supported by grants from the Fonds pour
la Formation des Chercheurs et l'Aide \`a la Recherche of the Government of
Qu\'ebec and from the Natural Science and Engineering Research Council of
Canada.  L.T. acknowledges the support of the Canadian Institute for Advanced
Research and the A.P. Sloan Foundation.

\begin{figure}
\caption{Transverse ultrasonic attenuation, $\alpha_{ab}$,
for propagation along $\hat a$
and polarization along $\hat b$. The low temperature attenuation shows no
curvature down to 130 mK.}
\label{Fig1}
\end{figure}

\begin{figure}
\caption{Transverse ultrasonic attenuation, $\alpha_{ac}$,
for propagation along $\hat a$
and polarization along $\hat c$.  The attenuation drops much more quickly than
$\alpha_{ab}$, reflecting the insensitivity of $\alpha_{ac}$ to
quasiparticles in the basal plane.}
\label{Fig2}
\end{figure}

\begin{figure}
\caption{
(a) Detail of $\alpha_{ab}$ in the A-phase.  After a drop at $T_c$,
the attenuation reaches a plateau before dropping sharply on
entering the B-phase.  This behavior reflects a larger quasiparticle
density in the A than in the B-phase.  The arrows designate $T_c^-$ and
$T_c^+$ as determined from longitudinal sound
velocity data[10].
(b) Detail of $\alpha_{ac}$ in the A-phase.  The onset of the B-phase is
evident as a small "bump" near $T_c^-$ = 435 mK.
}
\label{Fig3}
\end{figure}

\begin{figure}
\caption{
Transverse attenuation data for both polarizations normalized to the
value of the attenuation at $T_c^+$
and $T_c^-$ as a function of $T/T_c^{+,-}$.  This choice of
normalization allows us to compare the attenuation in the A and B phases
over the same (reduced) temperature range.  The A-phase shows an enhanced
in-plane attenuation vs. the B-phase while the out-of-plane polarization
data are roughly equal in the two phases.  The lines are guides to the
eye.}
\label{Fig4}
\end{figure}

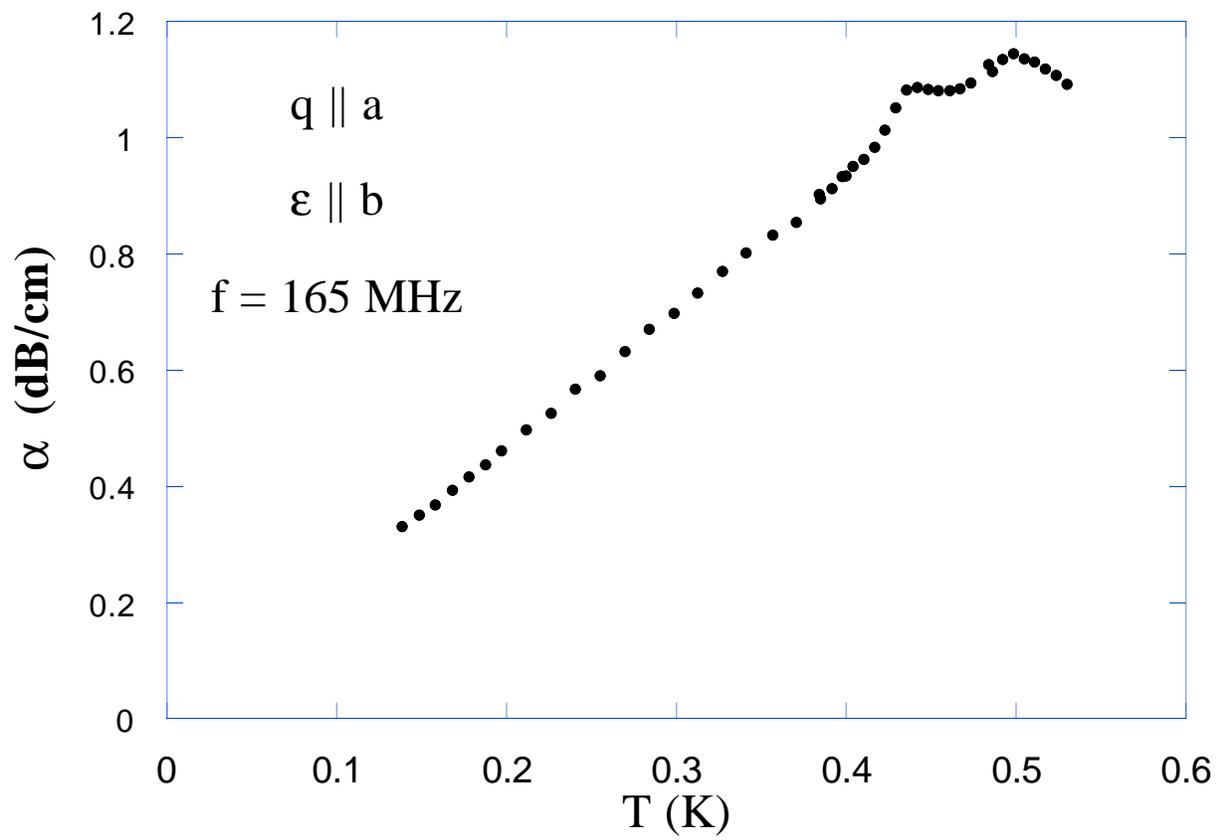
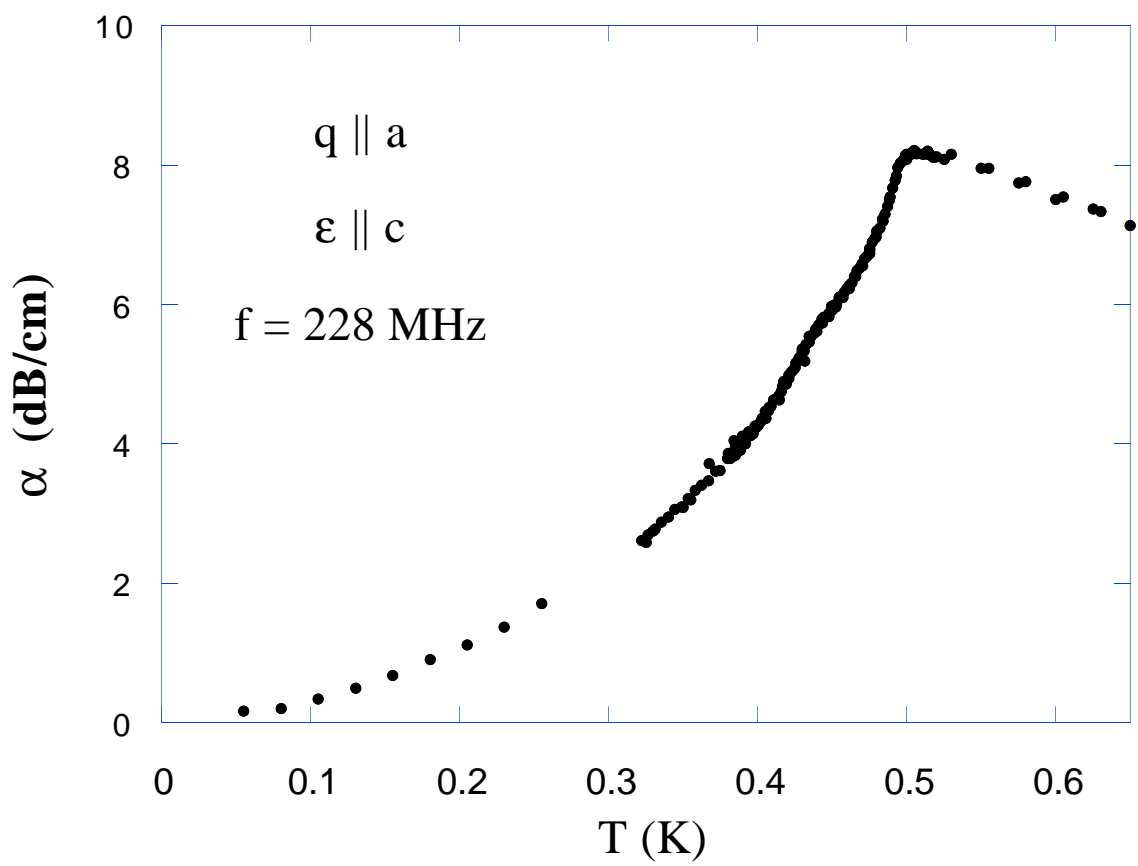
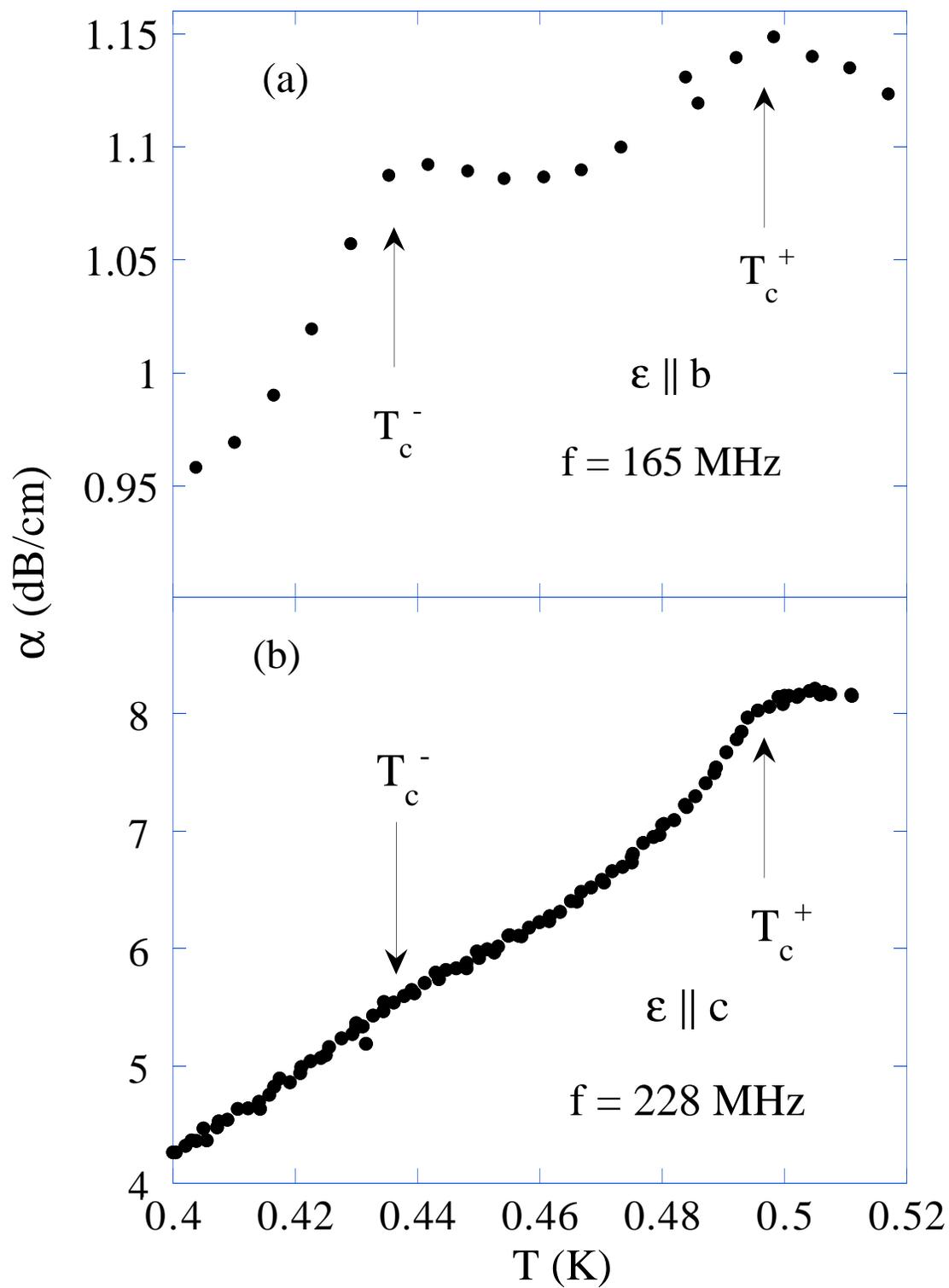
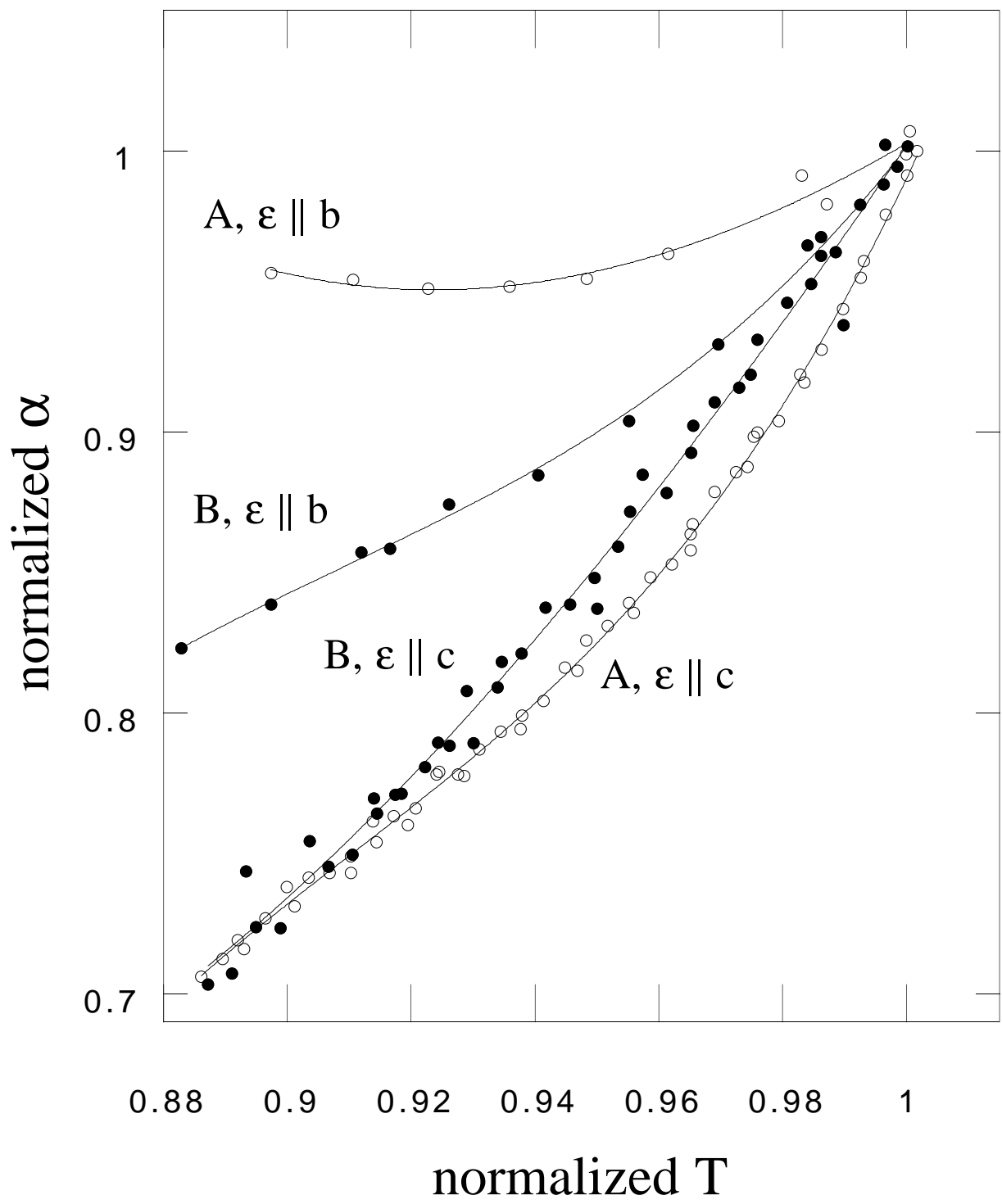
\end{document}